\documentclass[letterpaper, 10 pt, conference]{ieeeconf} 

\IEEEoverridecommandlockouts 

\usepackage{placeins}
\usepackage{graphics} 
\usepackage{epsfig} 
\usepackage{mathptmx} 
\usepackage{times} 
\usepackage{amsmath} 
\usepackage{amssymb} 
\usepackage{algorithm} 
\usepackage{verbatim} 
\usepackage{algpseudocode}
\usepackage{array}
\usepackage{epsfig}
\usepackage{amsmath}
\usepackage{float}
\usepackage{graphicx}
\usepackage{tabto}
\usepackage{cleveref}
\usepackage{subfigure}
\usepackage{wrapfig}
\usepackage{textcomp}
\usepackage{float}
\usepackage{booktabs}
\usepackage{graphicx}
\usepackage{url} 
\usepackage{multirow}
\usepackage{amsmath}
\usepackage{flexisym}
\usepackage{cite}
\usepackage{filecontents}
\usepackage{caption}
\usepackage{stackengine}

\title{\LARGE \bf
Rapid Quantification of White Matter Disconnection in the Human Brain
\\
}

\author{Abdelrahman Zayed$^{1}$, Yasser Iturria-Medina$^{2}$, Arno Villringer$^{3}$, Bernhard Sehm$^{4}$ and Christopher J. Steele$^{5}$
\thanks{*This research was funded by the Quebec Bio-imaging Network, Montreal, Quebec, Canada and Max Planck Institute for Human Cognitive and Brain Sciences, Leipzig, Germany.}
\thanks{$^{1}$Abdelrahman Zayed is with Department of Electrical and Computer Engineering
and PERFORM Centre, Concordia University, Montreal, Quebec, Canada
{\tt\small a\_zayed@encs.concordia.ca}}%
\thanks{$^{2}$Yasser Iturria-Medina is with Department of Neurology and Neurosurgery, McGill University, Montreal, Quebec, Canada
{\tt\small yasser.iturriamedina@mcgill.ca}}%
\thanks{$^{3}$Arno Villringer is the Director of the Department of Neurology, Max Planck Institute for Human Cognitive and Brain Sciences and Clinic for Cognitive Neurology, University of Leipzig, Leipzig, Germany
{\tt\small villringer@cbs.mpg.de}}%
\thanks{$^{4}$Bernhard Sehm is with Department of Neurology, Max Planck Institute for Human Cognitive and Brain Sciences and Clinic for Cognitive Neurology, University of Leipzig, Leipzig, Germany
{\tt\small sehm@cbs.mpg.de}}%
\thanks{$^{5}$Christopher J. Steele is with Department of Psychology, Concordia University, Montreal, Quebec, Canada
{\tt\small christopher.steele@concordia.ca}}%
}

\begin{document}

\maketitle
\thispagestyle{empty}
\pagestyle{empty}

\begin{abstract}

With an estimated five million new stroke survivors every year and a rapidly aging population suffering from hyperintensities and diseases of presumed vascular origin that affect white matter and contribute to cognitive decline, it is critical that we understand the impact of white matter damage on brain structure and behavior. Current techniques for assessing the impact of lesions consider only location, type, and extent, while ignoring how the affected region was connected to the rest of the brain. Regional brain function is a product of both local structure and its connectivity. Therefore, obtaining a map of white matter disconnection is a crucial step that could help us predict the behavioral deficits that patients exhibit. In the present work, we introduce a new practical method for computing lesion-based white matter disconnection maps that require only moderate computational resources. We achieve this by creating diffusion tractography models of the brains of healthy adults and assessing the connectivity between small regions. We then interrupt these connectivity models by projecting patients' lesions into them to compute predicted white matter disconnection. A quantified disconnection map can be computed for an individual patient in approximately 35 seconds using a single core CPU-based computation. In comparison, a similar quantification performed with other tools provided by MRtrix3 takes 5.47 minutes.

\end{abstract}

\section{INTRODUCTION}
The problem of relating the location of a lesion to its behavioral effect has been studied by many researchers \cite{charil2003statistical,mukherjee2005diffusion,niogi2008structural,pfefferbaum2005frontal,lampe2019lesion}. Bates \textit{et al.} \cite{bates2003voxel} developed a method called voxel-based lesion-symptom mapping (VLSM) which relates overlapping lesion locations to common behavioural deficits. This method does not require that a specific region of interest is identified, but does require overlap between the lesions. Depending on the type of clinical measure that is being targeted, different types of tests are performed to identify the regions responsible for a specific impairment. Rorden \textit{et al.} \cite{rorden2007improving} proposed further statistical enhancements that can provide more specific results than standard VLSM. In addition, Gleichgerrcht \textit{et al.} \cite{gleichgerrcht2017connectome} indicate that VLSM may be supplanted by the concept of connectome-based lesion-symptom mapping. This work, and our own, assumes that damage does not exist only in the lesion location, but also extends to all other regions that are connected to the lesion through the underlying white matter architecture. 
Kuceyeski \textit{et al.} \cite{kuceyeski2013network} proposed the first method for generating lesion-based white matter disconnection maps with data from diffusion-weighted magnetic resonance imaging (DW-MRI): the network modification (NeMo) tool. Damage is computed for the regions that are connected both directly (i.e. they lie inside the lesion volume) and indirectly (i.e. there is a white matter connection between the lesion and a given region). 

Although all of the mentioned methods have acceptable performance in identifying the regions that are responsible for behavioral deficits, they also have some drawbacks. Voxel-based methods such as \cite{bates2003voxel} and \cite{rorden2007improving} study the damage due to brain injury in a voxel-by-voxel manner, requiring lesion overlap within the population for statistical significance, and completely ignoring the fact that there are lesion effects that extend beyond the visible damage and throughout the brain, according to underlying white matter architecture. The method used in \cite{kuceyeski2013network} relies on a network of pre-defined anatomical regions and takes up to several hours to produce the disconnection map; thus restricting its general applicability and practical use.

In this paper, we introduce a new method based on human DW-MRI estimates of white matter connectivity that can take as little as 35 seconds to produce a lesion-based disconnection map. Given a patient lesion and known model of normative brain connectivity, we use the lesion's spatial location to interrupt connections and quantify the degree of disconnection between all regions of the brain. This method could be applied in a clinical context to predict behavioral deficits in individual patients and patient populations. Our method has been tested on both simulated and real data from a stroke patient, and provides rapid results that are similar to those produced with MRtrix3 \cite{mrtrix3}.

\label{sec:intro}

\section{METHOD}
In this section we describe our model and show how we can modify it to further improve the resolution of the disconnection map.
\subsection{Model description}
In order to be able to accurately compute the damage in every region in the brain as a result of having a lesion, we partition the brain volume into nodes, where each node is a cube of side length $l$, such that $l$ is a hyperparameter chosen by the user. Every node is further partitioned into even smaller units called voxels, such that a voxel is a tiny cube of dimensions $1$ $mm$ $\times$ $ 1$ $mm$ $\times$ $ 1$ $mm$. Hence, the node contains $l^{3}$ voxels. 

For any model of the brain, we use MRtrix3 (a set of tools for analyzing DW-MRI images) to generate probabilistic white matter connectivity between every pair of nodes. We then compute the two matrices that are used to calculate quantified disconnection maps. The first matrix is referred to as the connectivity matrix, which has the number of connections (probabilistic streamline counts) between any two nodes in the brain. The second matrix is the weights matrix, which describes the density of connections at each voxel. It is important to note that these two matrices need to be computed only once offline using MRtrix3, subsequent calculations use them to compute the disconnection map.

Given a lesion in any spatial location within the brain, we classify the affected nodes into two types: directly affected (nodes that lie inside the lesion volume) and indirectly affected (located outside the lesion volume but connected to directly affected nodes). In every node, we define a metric called damage, which is a number that goes from 0, when no connections are affected, to 1 when all connections from/to this node are affected. To compute the overall damage due to the lesion, we take the union of both the direct and indirect damage, as described in detail below.

\subsubsection{Computing the direct damage}
\begin{figure*}[h]
\captionsetup{justification=centering}
\vspace*{0.7em}
\begin{center}
\includegraphics[width=0.7\linewidth,trim=0 0 0 0,clip]{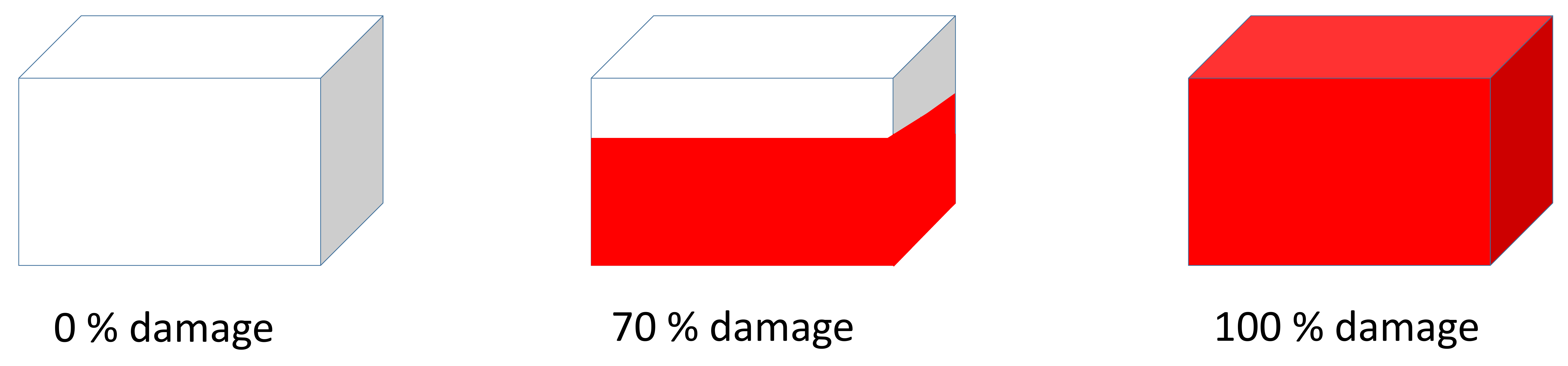}
\end{center}%
\centering
\caption{Computing the direct damage for different nodes.}%
\label{direct_damage}%
\end{figure*}

To compute the direct damage in a certain node that lies partially or completely inside the lesion volume, we calculate the ratio between the volume of the node inside the lesion to the total volume of the node. In other words, we calculate the ratio between the number of node voxels inside the lesion to the total number of node voxels as shown in Eq.\ref{direct_damage_eq}:
\begin{equation}
\label{direct_damage_eq}
d_{direct}[i]=\frac{N[i]}{l^3}
\end{equation}
such that: \begin{itemize}
\item $d_{direct}[i]$ is the direct damage at node $i$.
\item $N[i]$ is the number of voxels in node $i$ that lie inside the lesion volume.
\item $l^3$ is the total number of voxels in the node.
\end{itemize}
The direct damage is 1 for nodes that lie completely inside the lesion volume.~Fig.\ref{direct_damage} shows an example of how the direct damage is computed for different nodes.

\subsubsection{Computing the indirect damage}
To illustrate the concept of indirect damage, consider having two nodes A and B that are connected. Assume that node A has direct damage because it lies within the lesion volume. Node B, on the other hand, does not lie inside the lesion volume and therefore does not have direct damage. However, node B has indirect damage as a result of its connection to node A.

To calculate the damage that occurs to the nodes that do not lie inside the lesion volume (i.e. indirectly affected nodes), we need to know the number of connections they have with the directly affected nodes and with the rest of the nodes in the brain. This is computed using the connectivity matrix, where we infer indirect damage at node $i$ as follows:
\begin{equation}
\label{indirect_damage_eq}
d_{indirect}[i]=\frac{\sum_{j=1}^{Q} W_{ij} d_{direct}[j]}{\sum_{j=1}^{M} W_{ij}}
\end{equation}

where: \begin{itemize}
\item $d_{indirect}[i]$ refers to the indirect damage at node $i$.
\item $Q$ is the total number of directly affected nodes.
\item $W_{ij}$ is the number of connections between nodes $i$ and $j$.
\item $M$ is the total number of nodes that we have in the brain.
\end{itemize}

\begin{figure*}[h]
\begin{center}
\includegraphics[width=0.8\linewidth,trim=0 0 0 0,clip]{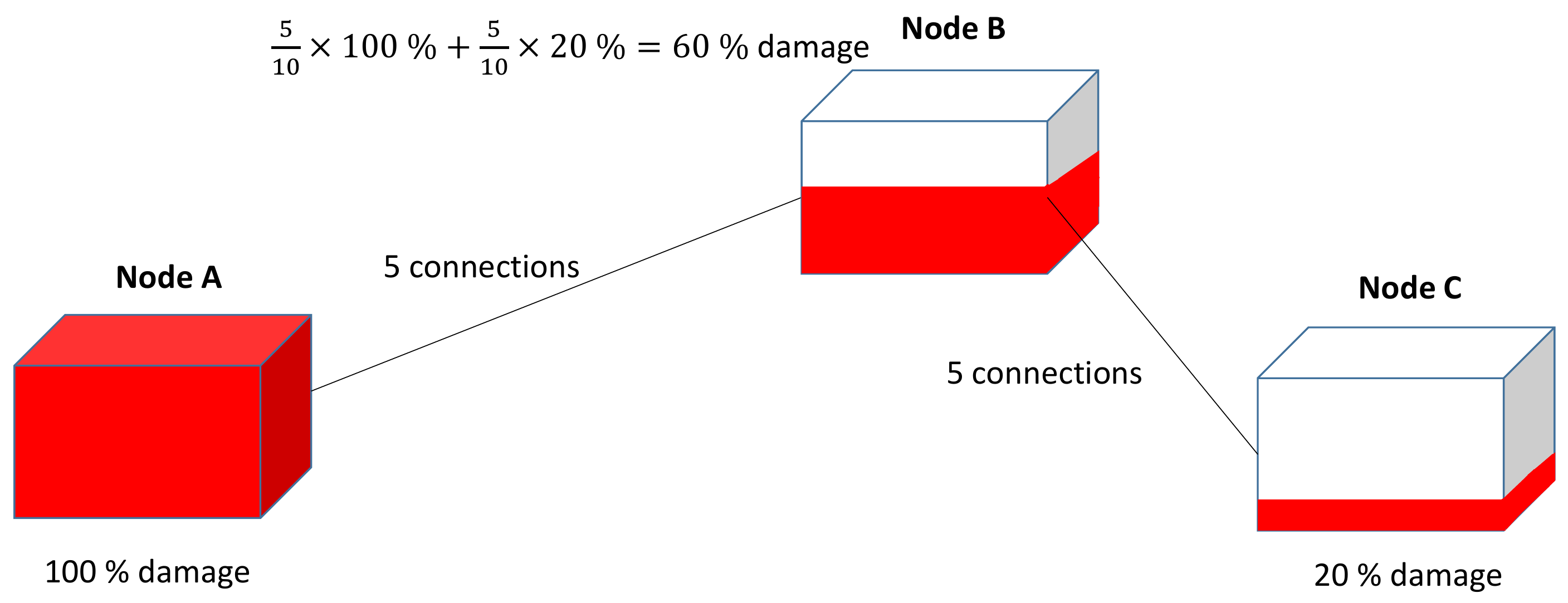}
\end{center}%
\centering
\caption{An example for computing the indirect damage for a certain node. Nodes A and C are directly affected as they lie partially or completely inside the lesion volume, whereas node B is indirectly affected as it is connected to them.}%
\label{indirect_damage}%
\end{figure*}

Fig.~\ref{indirect_damage} shows an example with three different nodes A, B and C. Assuming that nodes A and C are directly affected by a lesion, it is required to compute the indirect effect at node B. By applying Eq.~\ref{indirect_damage_eq}, we can deduce that the indirect damage at node B is 60\%.



\subsection{Increasing the resolution of the disconnection map}
The damage computed from Eq. \ref{direct_damage_eq} and \ref{indirect_damage_eq} is for the whole node. Since our nodes are of size $l$ $mm$ $\times$ $ l$ $mm$ $\times$ $ l$ $mm$, which is equivalent to $l^3$ voxels, this means that all of the $l^3$ voxels inside the node share the same damage. This is an acceptable approximation given the theoretical accuracy of probabilistic tractography for relatively small values of $l$, but as $l$ increases the resolution will degrade. The effective resolution of our approach can, however, be improved with minimal computational time. This is achieved by allowing different voxels inside a given node to have different damage. This yields an approximate $l^3$ times increase in the resolution and leads to a substantial improvement in the spatial accuracy of damage distribution within the node.

The idea depends on using the weights matrix that was previously generated by MRtrix3. Intuitively, voxels with high connectome density would have more damage compared to those with less connectome density inside the same node. Assuming we have node $i$ with a damage $d[i]$ (the union of $d_{direct}[i]$ and $d_{indirect}[i]$), instead of having this value in each of the $l^3$ voxels inside the node, $d[i]$ will change from one voxel to another by a modulation factor that is the ratio between the connectome density in a voxel to the average connectome density of its parent node. Therefore, if a voxel lying inside a certain node has twice the average connectome density of its parent node, its modulated damage will be twice its original damage.

\section{EXPERIMENTS AND RESULTS}
In this section, we compare the performance of our method on lesions from both simulated and real masks. We set $l$ to 5, rendering nodes of volume $125$ $mm^3$, for a total of 31,262 nodes inside the brain volume. We also show the effect of increasing the resolution of the disconnection map. After that, we will discuss the running time of our method, which is considered to be one of its main advantages.

\subsection{Results on simulated data}
\begin{figure*}[h]
\begin{center}
\includegraphics[width=1.0\linewidth,trim=0 0 0 0,clip]{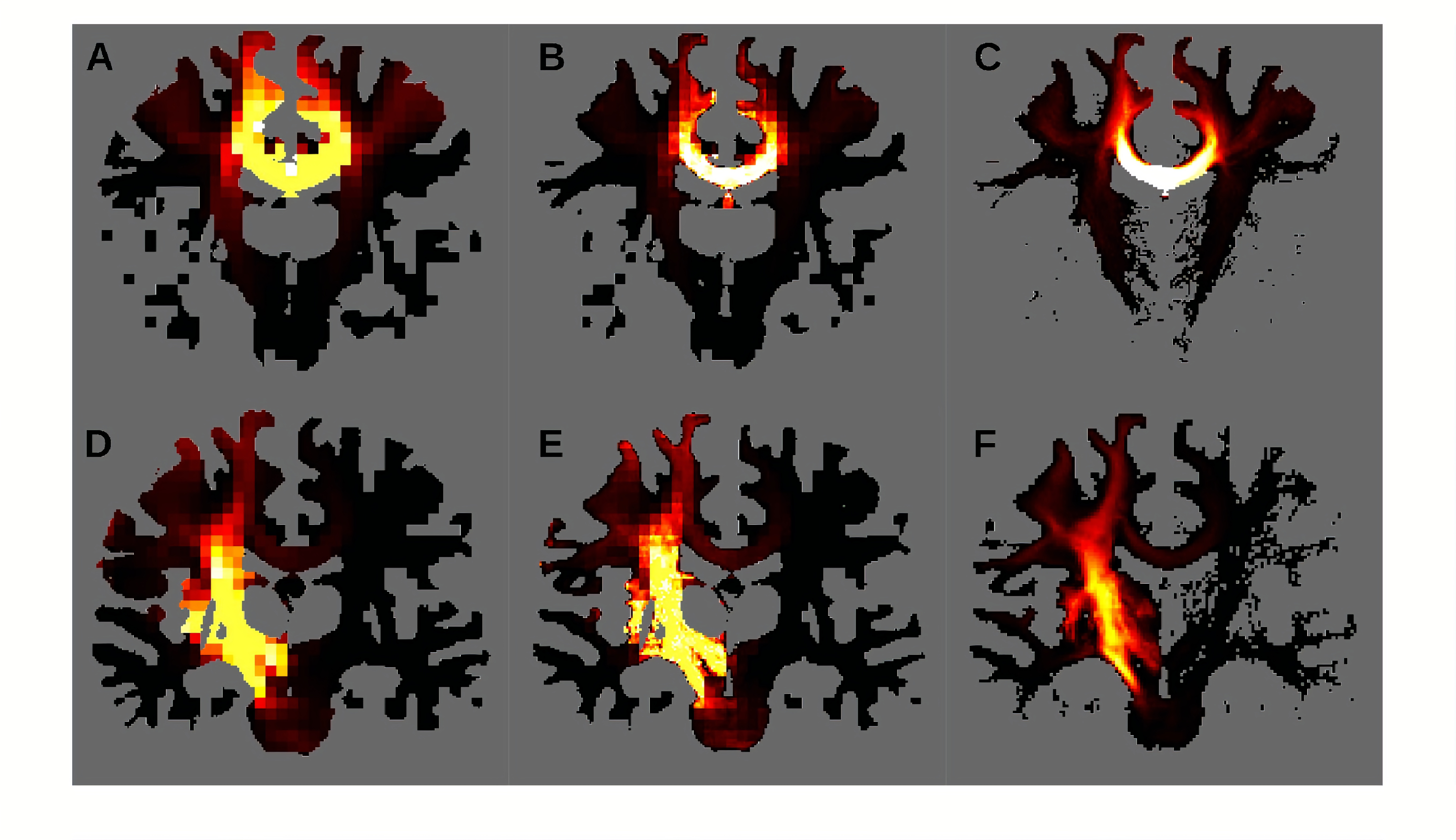}\vspace*{0.6em}
\stackunder[5pt]{\includegraphics[width=2in,height=.35in]{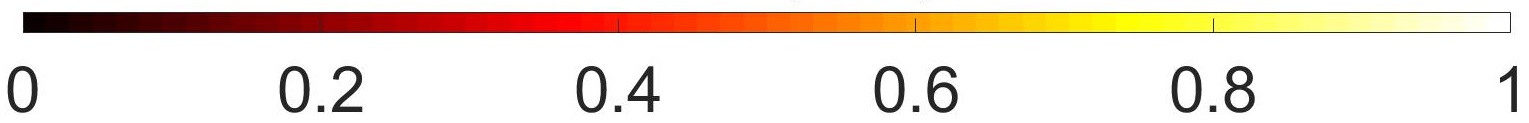}}{}
\end{center}%
\centering
\caption{Disconnection maps obtained using different methods. The first row shows the disconnection maps due to a simulated lesion in the corpus callosum, where A and B refer to the result of our method before and after modulation respectively, while C refers to the result of MRtrix3. The second row shows the results due to a real lesion from a stroke patient, where D and E refer to our method before and after modulation respectively, while F refers to MRtrix3.}%

\label{CC_S19_lesions}%
\end{figure*}

Fig.\ref{CC_S19_lesions} (A-C) shows a comparison between the disconnection map produced by both our method, before and after modulation to increase resolution, and MRtrix3 when tested on a lesion from a simulated mask in the corpus callosum. We can observe that the modification we introduced for the disconnection map substantially increased the resolution and brings our results closer to those of MRtrix3.

%

\subsection{Results on real data}
Fig.\ref{CC_S19_lesions} (D-F) shows a comparison between the disconnection map produced by both our method, before and after increasing the resolution, and MRtrix3 when tested on a lesion from a real mask obtained from a stroke patient. As observed before with the simulated data, our method gives a very close approximation to the actual disconnection map.

%

\subsection{Running time}
Our method takes an average of 35 seconds on an 8th generation 2.7 GHz Intel core i7 computer with a lesion that imparts direct damage to 33 nodes. All of our code is written in Python (v3.6). Computation time will scale with lesion volume. It is important to mention that the modification we introduced to increase the resolution of the disconnection map will only add an extra 1 second to the running time. MRtrix3, on the other hand, takes 5.47 minutes to produce a slightly higher resolution disconnection map.
\section{DISCUSSION AND FUTURE WORK}

In this paper, we present a rapid and practical method for quantifying the impact of white matter disconnection in individual patients that requires little computational and storage resources. Our method is more than 9 times faster than MRtrix3, which is currently one of the most widely used tools for the analyses of DW-MRI. We also showed how to increase the spatial resolution of our disconnection map by approximately 125 times, without adding significant overhead to the running time. The resulting maps of disconnection are patient-specific, quantitative, and therefore enable direct comparison between patients suffering from a variety of lesions in different locations and with different behavioral deficits. Our method leverages pre-computed connectivity and weights matrices to reduce storage space and decrease computation time. Each model is reduced from approximately 6 GB (raw probabilistic streamlines file used in the MRtrix3 comparison computations) to less than 300 MB (connectivity and weights matrices when $l$ = 5).

Our future work will focus on extending the current method to calculate the disconnection map from brain models computed from a variety of different brains, which would yield multiple disconnection maps. This would enable us to calculate an approximate population distribution for the damage in every node across different brain models. The median disconnection map should be a more accurate disconnection map, taking into consideration individual variability. Our method will also be embedded into a user-friendly tool that will be made publicly available. Further extension to our work would include relating the disconnection maps obtained to the behavioral effects on patients.

\addtolength{\textheight}{-12cm} 




\section*{ACKNOWLEDGMENT}

Patient lesion data was collected at the Max Planck Institute for Human Cognitive and Brain Sciences Neurology Clinic and provided by Dr. Bernhard Sehm with the help of Leila Gajiyeva. The simulated mask in the corpus callosum was provided by Dr. Christopher Steele. The model brain used to generate the probabilistic streamlines and connectivity matrices was obtained from the publicly available Human Connectome Project \cite{van2013wu}.

\bibliographystyle{IEEEtran}
\bibliography{strings,refs}

\begin{thebibliography}{10}
\providecommand{\url}[1]{#1}
\csname url@rmstyle\endcsname
\providecommand{\newblock}{\relax}
\providecommand{\bibinfo}[2]{#2}
\providecommand\BIBentrySTDinterwordspacing{\spaceskip=0pt\relax}
\providecommand\BIBentryALTinterwordstretchfactor{4}
\providecommand\BIBentryALTinterwordspacing{\spaceskip=\fontdimen2\font plus
\BIBentryALTinterwordstretchfactor\fontdimen3\font minus
  \fontdimen4\font\relax}
\providecommand\BIBforeignlanguage[2]{{%
\expandafter\ifx\csname l@#1\endcsname\relax
\typeout{** WARNING: IEEEtran.bst: No hyphenation pattern has been}%
\typeout{** loaded for the language `#1'. Using the pattern for}%
\typeout{** the default language instead.}%
\else
\language=\csname l@#1\endcsname
\fi
#2}}

\bibitem{charil2003statistical}
A.~Charil, A.~P. Zijdenbos, J.~Taylor, C.~Boelman, K.~J. Worsley, A.~C. Evans,
  and A.~Dagher, ``Statistical mapping analysis of lesion location and
  neurological disability in multiple sclerosis: application to 452 patient
  data sets,'' \emph{Neuroimage}, vol.~19, no.~3, pp. 532--544, 2003.

\bibitem{mukherjee2005diffusion}
P.~Mukherjee, ``Diffusion tensor imaging and fiber tractography in acute
  stroke,'' \emph{Neuroimaging Clinics}, vol.~15, no.~3, pp. 655--665, 2005.

\bibitem{niogi2008structural}
S.~N. Niogi, P.~Mukherjee, J.~Ghajar, C.~E. Johnson, R.~Kolster, H.~Lee,
  M.~Suh, R.~D. Zimmerman, G.~T. Manley, and B.~D. McCandliss, ``Structural
  dissociation of attentional control and memory in adults with and without
  mild traumatic brain injury,'' \emph{Brain}, vol. 131, no.~12, pp.
  3209--3221, 2008.

\bibitem{pfefferbaum2005frontal}
A.~Pfefferbaum, E.~Adalsteinsson, and E.~V. Sullivan, ``Frontal circuitry
  degradation marks healthy adult aging: evidence from diffusion tensor
  imaging,'' \emph{Neuroimage}, vol.~26, no.~3, pp. 891--899, 2005.

\bibitem{lampe2019lesion}
L.~Lampe, S.~Kharabian-Masouleh, J.~Kynast, K.~Arelin, C.~J. Steele,
  M.~L{\"o}ffler, A.~V. Witte, M.~L. Schroeter, A.~Villringer, and P.-L. Bazin,
  ``Lesion location matters: the relationships between white matter
  hyperintensities on cognition in the healthy elderly,'' \emph{Journal of
  Cerebral Blood Flow \& Metabolism}, vol.~39, no.~1, pp. 36--43, 2019.

\bibitem{bates2003voxel}
E.~Bates, S.~M. Wilson, A.~P. Saygin, F.~Dick, M.~I. Sereno, R.~T. Knight, and
  N.~F. Dronkers, ``Voxel-based lesion--symptom mapping,'' \emph{Nature
  neuroscience}, vol.~6, no.~5, p. 448, 2003.

\bibitem{rorden2007improving}
C.~Rorden, H.-O. Karnath, and L.~Bonilha, ``Improving lesion-symptom mapping,''
  \emph{Journal of cognitive neuroscience}, vol.~19, no.~7, pp. 1081--1088,
  2007.

\bibitem{gleichgerrcht2017connectome}
E.~Gleichgerrcht, J.~Fridriksson, C.~Rorden, and L.~Bonilha, ``Connectome-based
  lesion-symptom mapping (clsm): a novel approach to map neurological
  function,'' \emph{NeuroImage: Clinical}, vol.~16, pp. 461--467, 2017.

\bibitem{kuceyeski2013network}
A.~Kuceyeski, J.~Maruta, N.~Relkin, and A.~Raj, ``The network modification
  (nemo) tool: elucidating the effect of white matter integrity changes on
  cortical and subcortical structural connectivity,'' \emph{Brain
  connectivity}, vol.~3, no.~5, pp. 451--463, 2013.

\bibitem{mrtrix3}
J.-D. Tournier, R.~Smith, D.~Raffelt, R.~Tabbara, T.~Dhollander, M.~Pietsch,
  D.~Christiaens, B.~Jeurissen, C.-H. Yeh, and A.~Connelly, ``Mrtrix3: A fast,
  flexible and open software framework for medical image processing and
  visualisation,'' \emph{NeuroImage}, p. 116137, 2019.

\bibitem{van2013wu}
D.~C. Van~Essen, S.~M. Smith, D.~M. Barch, T.~E. Behrens, E.~Yacoub,
  K.~Ugurbil, W.-M.~H. Consortium, \emph{et~al.}, ``The wu-minn human
  connectome project: an overview,'' \emph{Neuroimage}, vol.~80, pp. 62--79,
  2013.

\end{thebibliography}
\end{document}